\documentclass[12pt,a4paper]{article}
\usepackage{amsmath}
\usepackage{amsfonts}
\usepackage{amssymb}

\title{On a possibility to combine the order effect with sequential reproducibility for quantum measurements}

\author{Irina Basieva\\
 Prokhorov General Physics Institute\\ Vavilov str. 38D, Moscow, Russian Federation\\
Andrei Khrennikov \\ 
International Center for Mathematical Modeling \\in Physics, Engineering, Economics, and Cognitive Science\\
Linnaeus University, V\"axj\"o-Kalmar, Sweden}

\begin{document}
\maketitle

\begin{abstract} In this paper we study the problem of a possibility to use quantum observables to 
describe a possible combination of  the order effect 
 with sequential reproducibility for quantum measurements. By the order effect we mean a 
dependence of probability distributions (of measurement results) on the order of measurements. We consider two types of the sequential  reproducibility:
adjacent reproducibility ($A-A$) and separated reproducibility($A-B-A$). The first one is 
reproducibility with probability 1 of a result of measurement of some observable $A$ measured twice, one $A$ measurement  after the other.
 The second one, $A-B-A$,  is  reproducibility  with probability 1 of a result of $A$ measurement when 
another quantum observable $B$ is measured between two $A$'s. Heuristically, it is clear that the second 
type of reproducibility is complementary to the order effect. We show that, surprisingly, for an important class of quantum observables given by 
positive operator valued measures (POVMs), this may not be the case. The order effect can coexist with a separated reproducibility as well as
adjacent  reproducibility for both observables $A$ and $B.$ However, the additional constraint in the form of separated reproducibility of the $B-A-B$ 
type makes this coexistence impossible. Mathematically, this paper is about the operator algebra for effects composing POVMs. The problem under consideration 
was motivated by attempts to apply the quantum formalism outside of physics, especially, in cognitive psychology and psychophysics. However, it is 
also important for foundations of quantum physics as a part of    the problem about the structure of sequential 
quantum measurements.
\end{abstract}
   
\section{Introduction}

Sequential measurements play an important role in quantum measurement theory, e.g., \cite{DL}--\cite{SQ}. In particular, sequential measurements of two 
incompatible observables $A$ and $B$ induce the probability order effect which gives the probabilistic representation of the 
principle of complementarity, in the form of nonexistence of the joint probability distribution of such observables. In other words,
sequential probability distributions $p_{A-B}(x,y)$ and $p_{B-A}(y,x)$ do not coincide. 

We remark that if $A$ and $B$ are quantum measurements of 
the von Neumann-L\"uders type (measurements of the first kind), i.e., they are mathematically represented by Hermitian 
operators and the state transformers (quantum operations)\footnote{The terminology ``state transformer'' \cite{BL} is may be old fashioned. 
In modern literature \cite{Jaeger} the terms of quantum operation and quantum channel are typically in the use. The latter notions are motivated 
by quantum information theory \cite{Jaeger}. However, the problems handled in this paper have no direct relation to quantum information theory. They are 
motivated by applications of the quantum calculus of probabilities outside of physics, see the last part of this 
section for details. In such applications
the terminology ``state transformer'' seems to be more appropriate.}
are given by the  orthogonal projectors corresponding to the eigenvalues\footnote{In the mathematical model, see, e.g., \cite{DL}--\cite{SQ}, 
quantum measurement is represented with the aid of two structures: a quantum observable (Hermitian operator or more generally POVM) and a state transformer.
}, then they have the property of  
{\it adjacent sequential reproducibility}: for $A-A$ and $B-B$ measurements the values observed in the first measurement are reproduced 
(with probability 1) in the second measurement. Thus, for  quantum measurements of the first kind, 
the condition of adjacent reproducibility is a redundant constraint. However, for quantum measurements of the second kind,
i.e., those which cannot be represented as measurements of the first kind, this constraint is nontrivial. 

We remark that, for quantum measurements of the second kind neither 
an observable nor a state transformer have to be represented by orthogonal projectors. In general, for measurements of the 
second kind, observables are represented by POVMs. We also remark that a quantum measurement with an observervable given by a Hermitian operator 
can also be of the second kind: if the corresponding state transformers are not given by orthogonal projectors. The latter class of quantum 
measurements will play an important role in problems under the study in this paper.

For measurements of the first kind, the order effect can be approached only 
in case of incompatibility: commutativity of the observables $([A,B]=0)$ implies the existence 
of the joint probability distribution serving both sequential measurements, $A-B$ and $B-A.$  
Intuitively, in the case of incompatible observables a measurement of the observable $B$ after a preceding measurement of the observable $A$ (with the result $A=x)$ modifies crucially the post-measurement state $\psi_A$ generated as the result of the $A$ measurement. Thus, in the post-measurement state after the sequence 
of measurements $A-B$, the information about the value $A=x$ is at least partially washed out and one cannot expect that in
the sequence of measurements $A-B-A$ the value $A=x$ would be reproduced (obtained in the second $A$ measurement) with probability 1. Thus, one would expect that
the quantum order effect cannot be combined with both the adjacent and separated reproducibility. 
In paper \cite{BKDB} it was shown that, 
for measurements of the first kind, this is really the case.  

The problem of an extension of this result 
to  measurements of the second kind is mathematically more complicated. Some special class of  measurements of the second kind was considered 
in the aforementioned paper and it was shown that the combination of $A-A, B-B,$ and  $A-B-A$ reproducibilities implies disappearance 
of the order effect. The class of  measurements of the second kind considered in  \cite{BKDB} is characterized by the following conditions:

\medskip
${\bf C0}$ Observables are represented by POVMs.

\medskip

${\bf C1}$ The state transformer corresponding to an effect $E$ has the form 
\begin{equation}
\label{hum}
\psi \to \frac{M\psi}{\Vert M \psi \Vert},
\end{equation}
where 
\begin{equation}
\label{hh}
E=M^*M
\end{equation}
is some representation of the effect.

\medskip

${\bf C2}$ If $E=P$ is an orthogonal projector, then its state transformer is given by this projector
\begin{equation}
\label{hum1}
\psi \to \frac{P\psi}{\Vert P \psi \Vert}.
\end{equation}

The condition ${\bf C1},$ although it restricts the class of possible state transformers, is very natural,
because the majority of state transformers used in applications really have this simple form, (\ref{hum}). 
By the condition ${\bf C2}$ the class of measurements corresponding to the effects given by projectors  coincides 
with the class of measurements of the von Neumann-L\"uders type (measurements of the first kind). Although this condition is also quite natural, 
it restricts essentially the class of state transformers. In this paper we want to proceed with the class of quantum measurements
described solely by the conditions ${\bf C0}, {\bf C1}.$ Thus, even for an effect given by an orthogonal projector, in general  
the corresponding  state transformer does not coincide with this
projector. 

Surprisingly, this extension of the class of quantum measurements makes the situation essentially more complicated (and interesting).   
The order effect can coexist with  separated reproducibility  of $A-B-A$ type as well as
adjacent  reproducibility for both observables $A$ and $B.$ However, the additional constraint in the form of separated reproducibility of the $B-A-B$ 
type makes this coexistence impossible.

Finally, we remark that this study (as well as the preceding study \cite{BKDB}) was motivated by applications of the quantum formalism outside of physics, namely,
in cognitive psychology and psychophysics, see, e.g., \cite{40}-\cite{C1} for introduction. In such applications the mathematical formalism of 
quantum mechanics is treated as an operational formalism, see, e.g., M. D' Ariano \cite{DARIANO},  for handling probabilistic data collected in aforementioned domains of research. This formalism 
works well for a wide class of psycho-effects, e.g., the disjunction effect and the order effect. In particular, to represent 
these two effects one can use 
the von Neumann-L\"uders measurements. As is well known, such observables satisfy the condition of  
adjacent reproducibility, i.e., $A-A$ reproducibility, which is natural for the majority of experiments of cognitive psychology.
 However, as was pointed out in \cite{BKDB}, in psychophysics one can find 
the experimental situations violating adjacent reproducibility.   Therefore one has to proceed with generalized observables given by 
POVMs.\footnote{Observables of this class appeared in the applications of the quantum mathematical formalism to humanities even by 
another reason \cite{UB}:
in experiments performed, e.g., in cognitive psychology,
 the matrices of transition probabilities for observables with non-degenerate spectra are typically not double-stochastic. However, for 
the von Neumann-L\"uders observables, they should be double-stochastic.} And this was the natural step in development of quantum-like modeling 
in humanities. 

As was first understood by the authors of \cite{BKDB}, there was a hidden pitfall in the rapidly increasing stream of applications of 
the quantum formalism to humanities, namely, the problem of separated reproducibility. As was pointed out in \cite{BKDB}, some experimental 
contexts of cognitive psychology are characterized by separated reproducibility effect in combination with the order effect. For example, political opinion 
polls often demonstrate order effects, but here, e.g., by replying ``yes'' to the first question $A,$ and ``no'' to the second question $B$
a respondent is typically  ``firm in her preferences'' expressed  in the form of the $A$-yes, so she will practically definitely say ``yes''
again if asked the $A$-question again, i.e., in the $A-B-A$ experiment. We also remark that in known and thinkable experimental contexts
in cognitive psychology, the same should also happen for the $B-A-B$ experiment. As was shown in \cite{BKDB}, it is impossible 
to use the measurements of the first kind to describe such a situation. The following natural question was posed by the authors of \cite{BKDB}:
{\it Is the operational quantum formalism powerful enough to cover all possible experimental contexts arising in humanities?} 
In this paper, we continue to work to find an answer to this question. And for the moment, the answer is negative. However, although this paper 
covers  a wide and natural class of state transformers, it is still not the most general. There is  still a ``loophole in the proof'' that   
the power of quantum methods in humanities is restricted. We remark that in principle there is no reason to expect that 
the quantum-like operational formalism 
serving humanities would coincide with the quantum physical formalism.  It may happen that novel quantum-like models would be explored, cf. \cite{UB}. 
     
Although the problem of combination of the order effect with the two types of reproducibility was motivated by applications of the quantum formalism 
outside of physics, mainly cognitive psychology, it is also important for quantum foundations as a part of the problem about the structure of sequential 
quantum measurements, cf. \cite{SQ}. There is also an important experimental dimension: {\it Can one find physical measurements exhibiting combination 
of the order effect with adjacent and separated reproducibilities?}

In this paper we consider only the case of finite dimensional state spaces. (Only such state spaces are used up to now in applications to humanities.)
The situation in the infinite-dimensional case is very different, see, e.g., Proposition 8,  and this case has to be studied separately.

\section{The basic consequence of adjacent reproducibility}

In this section we show that the class of effects  corresponding to measurements determined by 
the conditions ${\bf C0}, {\bf C1},$  and satisfying the condition of adjacent reproducibility 
coincides with the class of orthogonal projectors. Hence, the class the corresponding POVMs coincides with the class 
of von Neumann-L\"uders observables. But, in general, the state transformers are not reduced to projectors.  Here the basic result 
is presented in Theorem 1. (This statement was formulated in \cite{BKDB}. However, its proof contained a loophole which could not be closed.)

\medskip

{\bf Theorem 1.} {\it Let $E$ be a Hermitian operator such that $0 \leq E \leq I$ (an effect) and let $\langle E\phi, \phi\rangle=1$
for some pure state $\phi.$ Then this $\phi$ is an eigenvector of $E$ with the eigenvalue $\lambda=1.$}

{\bf Proof.} Consider in $H$ a basis $(e_j)$ consisting of eigenvectors of $E,$ i.e., 
$E e_j= \lambda_j e_j.$ Here $0\leq \lambda_j \leq 1.$ We have $E \phi= \sum_j \lambda_j \phi_j e_j,$ 
where $\phi_j = \langle \phi, e_j \rangle.$ We have $ \langle \phi, \phi \rangle = \sum_j \vert \phi_j \vert^2=1.$ We also have
$\langle E\phi, \phi \rangle = \sum_j \lambda_j \vert \phi_j \vert^2=1.$ Set $O_{\phi}=\{j: \phi_j \not=0\}$ (this set depends on $E).$ Then 
we have the system of two quadratic equalities:
\begin{equation}
\label{tttt}
 \sum_{j\in O_{\phi}}  \vert \phi_j \vert^2=1,  
\end{equation}
\begin{equation}
\label{tttt1}
\sum_{j\in O_{\phi}} \lambda_j \vert \phi_j \vert^2=1.
\end{equation}
If at least one $\lambda_j$ in (\ref{tttt1}) is strictly less than 1, then we come to contradiction with  (\ref{tttt1}).
Thus, in (\ref{tttt1}) all $\lambda_j=1,$ i.e., if $j \in O_{\phi},$ then $\lambda_j=1.$ 
Hence $ E\phi = \sum_{j \in O_{\phi}} \phi_j e_j = \phi.$  

\medskip

{\bf Corollary 1.} {\it For an effect $E$ and  the corresponding state transformer of the type ${\bf C1},$ 
the condition of adjacent reproducibility is equivalent to the operator 
equality:}
\begin{equation}
\label{tmu}
E M =M  .
\end{equation}

\medskip

{\bf Theorem 2.} {\it For an effect  $E$ and  the corresponding state transformer of the type ${\bf C1},$   the operator equality (\ref{tmu}) 
is equivalent to the condition ``$E$ is an orthogonal projector''.} 

{\bf Proof.} Consider the representation of all operators by matrices in the orthonormal basis $(e_j, j=1,...,n)$ 
consisting of eigenvectors of the operator $E: E=(E_{ij})=\rm{diag} (\lambda_1,..., \lambda_n), M=(m_{ij}).$ 
Suppose that this operator has an eigenvalue, e.g.,  $\lambda=\lambda_1,$ which is different from 0 and 1. 
Consider the vector given by the first row in $M, w= (m_{11}, ...., m_{1n}).$ Then the equality (\ref{tmu}) implies 
that $\lambda_1 w=w,$ and, hence, $w=0.$ Thus the first row of $M$ consists of zeros (hence, the first column of $M^*$ also
consists  of zeros). 

Consider the vectors $v_1=(m_{21},..., m_{n1}),..., v_n=(m_{2n},..., m_{nn}),$ the columns of $M$ without the first
zero element. Since $E=M^* M,$ its matrix elements can be represented as 
\begin{equation}
\label{tmu7}
E_{ij}= \langle v_i, v_j\rangle.
\end{equation}
Since the off-diagonal elements in $E$ are zeros, we have:
$ v_1 \perp v_j, j\not=1, ..., v_n \perp v_j, j\not=n.$  Consider the first condition of orthogonality. Here 
 a vector in the $n-1$ dimensional space, namely, $v_1,$ is orthogonal to $n-1$ vectors. There are two possibilities: either $v_1=0$ 
or vectors $v_2,...,v_n$ are linearly dependent. In the first case (\ref{tmu7}) implies that 
$\lambda_1= \langle v_1, v_1\rangle=0.$ 
Suppose now that $v_1\not=0,$ and,  e.g., for $k=2,$ $v_2= \sum_{j\not=1, 2} c_j v_j.$ But $v_2 \perp v_j, j \not=2,$ hence $v_2=0.$ This means that 
vectors $v_3,..., v_n$ are linearly dependent, e.g., $v_3= \sum_{j\not=1, 2, 3} c_j v_j.$ In this way we get that $v_3$ is also zero
and so on. Finally, we get that all vectors $v_2,..., v_{n-1}$ equal to zero. The vectors $v_1$ and $v_2$ have to be orthogonal. 

Now let us take again the relation $E M =M$ into account.  
Consider vectors $w_j= (m_{j1}, ...., m_{jn}), j=1,...,n$ (the rows of the matrix $M$); so the vector $w$ used above coincides with $w_1.$  
We have $\lambda_2 w_2=w_2,$ but $\lambda_2=\langle v_2, v_2\rangle=0,$ so $w_2=0.$ In the same way we get that $w_3,..., w_{n-1} =0.$
Thus in the vectors $v_1$ and $v_2$ only the last coordinate can be nonzero, but such vectors can be orthogonal only if one of them is zero.
The worst case would be $v_1\not=0,$ but $v_n=0.$ But then $\lambda_n=0$ and hence $w_n=0,$ so the last coordinate in $v_1$ also has to be zero.

\section{The structure of state transformers for effects of the projection type}    

In the previous section we showed that, for measurements satisfying the condition of adjacent reproducibility, the effects compositing 
observables, POVMs, are, in fact, orthogonal projectors. Now we plan to describe the structure of the corresponding state transformer 
operators, see ${\bf C1}.$  We shall show that they are simply the compositions of the projectors-effects with unitary operators. This mathematical fact 
will play the crucial role in our further studies. 

We start with two simple lemmas about unitary operators which will be useful in the further considerations

\medskip

{\bf Lemma 1.} {\it Let $U: H\to H$ be a unitary operator and let $X$ be its invariant subspace, i.e., 
$U X \subset X.$ Then the orthogonal complement of $X$ is also invariant subspace of $U,$ i.e., 
$U X^\perp \subset X^\perp.$}

{\bf Proof.} Since the kernel of a unitary operator is zero, i.e., $Ux=0$ iff $x=0,$ dimensions 
of the spaces $X$ and $UX$ are equal. Thus, for  invariant subspace, we have that $UX=X.$ Hence, any 
$x\in X,$ can be represented as $x= U x_0, x_0 \in X.$ Take any $y \in X^\perp.$ Then $\langle U y, x\rangle= 
\langle U y, U x_0\rangle= \langle  y,  x_0\rangle=0.$

\medskip
{\bf Corollary 2.} {\it For any invariant subspace $X,$ the unitary operator $U$ can be decomposed as $U=\rm{diag} (V, W),$ where 
$V:X\to X$ and $ W:X^\perp \to X^\perp$ are unitary operators.}   

\medskip
     
Consider representation of an orthogonal projector $P$ in the form used in ${\bf C1}$ to define the 
corresponding  state transformer:
\begin{equation}
\label{SVD0}
P= M^* M.
\end{equation}
 Let us fix some basis and represent all operators by matrices. 
It is convenient to select a basis in which $P$ is diagonal $P =\rm{diag} (1,...,1, 0,...,0).$ 
By the singular value decomposition theorem any matrix $M$ can be represented in the form   
\begin{equation}
\label{SVD}
M= W \Sigma V^*,
\end{equation}
where the matrices $W$ and $V$ are unitary and $\Sigma$ is a diagonal matrix such that its elements are square roots of 
the eigenvalues of the matrix    $M^* M.$ Hence, in the case $M^* M=P$ the matrix $\Sigma$ coincides with $P.$ Thus,  
in the representation (\ref{SVD0}) we can always select
\begin{equation}
\label{SVD0a}
M = W P V^*.
\end{equation}
Moreover, we have  $P= M^* M= V P V^*$ or $ VP=PV$ and $V^*P =P V^*.$ Thus the representation (\ref{SVD0a}) can be written 
as  
\begin{equation}
\label{SVD0b}
M = W V^* P.
\end{equation}
We remark that the composition of two unitary operators is a unitary operator. 
Thus we obtained the following mathematically simple (but very useful for our further studies) result  

\medskip

{\bf Lemma 2.} {\it  All representations of the  projector $P$ in the form  
(\ref{SVD0}) are given by operators having the form:}
\begin{equation}
\label{SVD0d}
M = U P,
\end{equation}
where $U$ is a unitary operator.
 
\section{Main results on combination of adjacent and separated reproducibilities with the order effect}
 
Let $M= U P,$   
where $P$ is a projector and $U$  is a unitary operator. Set $H_P= P H.$   
Set $A= M^* M= P.$  So, the observable $A$  
is given by  a projector (Hermitian!), but the corresponding state transformer
is not of the L\"uders type (not measurement of the first kind):
\begin{equation}
\label{ST}
\psi \to \phi_A=\frac{U P\psi}{\Vert U P\psi\Vert}.
\end{equation}

{\bf Proposition 1.} {\it Observable $A=P$ with the state transformer (\ref{ST})  
has the property of $A-A$ repeatability iff}
\begin{equation}
\label{ST0}
PUP=UP
\end{equation}

{\bf Proof.} $$p_{A-A}=\langle P \phi_A, \phi_A\rangle=1.$$
Then, as we know, $$
P \phi_A= \phi_A.
$$ 
Thus, $P M \psi= M\psi$ or $PUP=UP.$

\medskip

{\bf Corollary 2.} {\it Observable $A=P$ with the state transformer (\ref{ST})  
has the property of $A-A$ repeatability iff the subspace $H_P$ is invariant 
with respect to the action of the unitary operator $U.$  }   
 
\medskip
 
Now we consider two observables: 

Let $M= U_1 P_1$ and $N= U_2 P_2,$  
where $P_j, j=1,2,$ are projectors and $U_j$  are unitary operators. Set $H_j= P_j H$ and $H_{12}= H_1 \cap H_2.$

Set $A= M^* M= P_1U_1^*U_1 P_1=P_1$ and $B= N^* N= P_2U_2^*U_2 P_2=P_2.$ So, the observables $A$ and $B$ 
are given by  projectors $P_j$ (Hermitian!), but the corresponding state transformers
are not of the L\"uders type (these are not measurements of the first kind):

\medskip

{\bf Proposition 2.} {\it A pair of observables $A$ and $B$ exhibits the order effect iff}
\begin{equation}
\label{ST1}
P_1U_1^* P_2 U_1P_1 \not = P_2 U_2^* P_1 U_2P_2
\end{equation}

{\bf Proof.} For the $A-B$ measurement sequence, we have:  
$$
p_{A-B}= \Vert \phi_A \Vert ^2 \langle P_2 \phi_A, \phi_A \rangle = \langle P_2 U_1P_1 \psi, U_1P_1 \psi \rangle=
 \langle P_1U_1^* P_2 U_1P_1 \psi,  \psi \rangle;
$$
in the same way, for  the $B-A$ measurement sequence, we obtain:  
$$
p_{B-A}= 
 \langle P_2U_2^* P_1 U_2P_2 \psi,  \psi \rangle;
$$
hence, these sequential measurements can give different results iff
(\ref{ST1}) holds true.

\medskip

{\bf Proposition 3.} {\it A pair of observables $A$ and $B$ exhibits the $A-B-A$ repeatability iff}
\begin{equation}
\label{ST2}
P_1N M= N M, \mbox{or}\; P_1 U_2 P_2U_1 P_1 = U_2 P_2U_1 P_1. 
\end{equation}
{\bf Proof.} For the $A-B-A$ measurement sequence, we have :  
$$
p_{A-B-A}= \langle P_1 \phi_{A-B}, \phi_{A-B} \rangle, 
$$
where 
$$
\phi_{A-B}= \frac{NM \psi}{\Vert NM \psi \Vert}.
$$
The condition $p_{A-B-A}=1$ is equivalent to the condition $P_1  \phi_{A-B} =\phi_{A-B},$
or (\ref{ST2}).

\medskip

Thus, we are looking for observables satisfying conditions (\ref{ST0})--(\ref{ST2}).

For simplicity, set $U_2=I,$ i.e., $B$ is measurement of the first kind.
Then we have the system of relations
\begin{equation}
\label{ST0a}
P_1U_1P_1=U_1 P_1 \; \mbox{or equivalently}\; U_1: H_1 \to H_1;
\end{equation}
\begin{equation}
\label{ST1a}
P_1 U_1^* P_2 U_1P_1 \not = P_2 P_1 P_2;
\end{equation}
\begin{equation}
\label{ST2a}
P_1 P_2 U_1 P_1 = P_2 U_1 P_1. 
\end{equation}
       
We shall show by an example that these conditions can be jointly satisfied. 
This statement can be reformulated in the form of the following proposition 
playing an important role in applications to cognition \cite{BKDB}-\cite{C1}.

\medskip

{\bf Proposition 4.} {\it There exist quantum observables $A$ and $B$ 
generating the order effect and satisfying the conditions of $A-A, B-B,$ and 
$A-B-A$ repeatability.}  

\medskip       
       
{\bf Example.} Consider four dimensional Hilbert space with some orthonormal basis $(e_1,e_2,e_3, e_4).$
Let $H_{1}$ and $H_{2}$ have bases $(e_1, e_2, e_3)$ and $(e_1,e_2, e_4),$ respectively, and let $P_1$ and 
$P_2$ be projectors on these subspaces. It is crucial that $H_{12} = H_1 \cap H_2$ (with the basis $(e_1, e_2)$) 
is nontrivial and at least two dimensional.

Let $U=(u_{ij})$ be a unitary operator acting in $H_1.$ It is extended on $H$ as $U_1=\rm{diag} (U,I),$ where $I$ is the unit operator 
in $H_1^\perp$ and $(e_4)$ is the basis in the latter space.  Thus the 
condition (\ref{ST0a}) is satisfied. 

We check now condition (\ref{ST2a}). For any $\psi \in H,$ the vector $U_1 P_1 \psi \in H_1.$ But $P_2: H_1 \to H_{12}.$ Hence, $P_2 U_1 P_1 \psi \in H_{12}$ and 
action onto this vector by $P_1$ cannot change it. 

Finally, we show that the condition (\ref{ST1a}) holds. The crucial point is that the right-hand side does not contain 
the unitary operator $U_1,$ but the left-hand side contains it. The tricky point is that the left-hand side  
contains both $U_1$ and it inverse $U_1^*.$ Thus, in principle, they can compensate actions of each other (cf. with the proof 
of Proposition 7). We shall see 
that this is not the case.  

Let $\psi= \sum_{j=1}^4 x_j e_j.$ Then $\phi= P_2 P_1 P_2 \psi \in H_{12}$ and $\phi= x_1 e_1+ x_2 e_2.$   

For the other side of (\ref{ST1a}), we have: $U_1P_1 \psi= \sum_{j=1}^3 y_j e_j,$ where $y_j= \sum_{i=1}^3 u_{ji} x_i.$ Now consider the next step:
$P_2 U_1P_1 \psi=  y_1 e_1 + y_2 e_2 .$ Then 
$$
\phi^\prime= U_1^* P_2 U_1P_1 \psi=\sum_{k=1}^3 z_k e_k,$$ where $z_k = \sum_{j=1}^2 \bar{u}_{jk} y_j= \sum_{j=1}^2 \sum_{i=1}^3 \bar{u}_{jk} u_{ji} x_i=
\sum_{i=1}^3  (\sum_{j=1}^2 \bar{u}_{jk} u_{ji}) x_i .$     
Finally, since $\phi^\prime \in H_{1},$ it cannot be changed by the action of $P_1.$    

We remark that, for a unitary matrix $U,$ 
$\sum_{j=1}^3 \bar{u}_{jk} u_{ji}= \delta_{ki}.$ If the subspace $H_{12}$ is also invariant 
for $U_1$ (and, hence, for its block $U$), i.e., e.g., $U=\rm{diag} (W, 1),$ where $W$ is a unitary 
operator in $H_{12},$ then  $\sum_{j=1}^2 \bar{u}_{jk} u_{ji}= \delta_{ki}.$ Hence, in this case $\phi^\prime=  x_1 e_1+ x_2 e_2= \phi.$
However, if $H_{12}$ is not invariant, then in general $\phi^\prime \not= \phi.$
          
\medskip

Now about the possibility to satisfy both $A-B-A$ and $B-A-B$ repeatability in combination with the order effect. As we have seen in the above 
example, for a nontrivial order effect it is crucial that the unitary operator does not leave the space $H_{12} = H_1 \cap H_2$ invariant. Now we shall take this 
into account in the general case. We turn again to the general case of two unitary  operators $U_1, U_2.$  We collect the list of conditions: 
\begin{enumerate}
\item $P_1 U_1^* P_2 U_1 P_1 \not = P_2 U_2^* P_1 U_2 P_2;$
\item $P_1 U_2 P_2 U_1 P_1 = U_2 P_2 U_1 P_1;$
\item $P_2 U_1 P_1 U_2 P_2 = U_1 P_1 U_2 P_2.$ 
\end{enumerate}

\medskip

{\bf Proposition 5.} {\it  The  combination of $A-A, B-B,$ and $A-B-A$ conditions  of sequential repeatability implies that} 
\begin{equation}
\label{Z1}
P_2 H_1= H_{12}
\end{equation}
 and 
\begin{equation}
\label{Z2}
U_2 H_{12} =H_{12}.
\end{equation}  

{\bf Proof.}  We start with remarking that $P_1 H=H_1$ and $U_1 P_1 H= H_1$ as well (Corollary 2 for the condition $A-A$). Then, we have that 
$P_2 U_1 P_1 H= P_2 H_1 \subset H_2$ and, hence, $U_2 P_2 U_1 P_1 H= U_2 (P_2 H_1) \subset H_2$ (Corollary 2 for the condition $B-B$). 
However, the second equality in the above list 
of conditions implies that $U_2 P_2 U_1 P_1 H  = U_2 (P_2 H_1) \subset H_1$ as well.  Thus, $U_2 (P_2 H_1) \subset H_1 \cap H_2= H_{12}.$ Now 
we play by using finite dimensionality of the model. We remark that $H_{12} \subset H_1$ and, hence, $H_{12} \subset P_2 H_1.$ We remark 
that the dimensions of the subspaces $P_2 H_1$ and $  U_2 P_2 H_1$ coincide. Since   $H_{12} \subset P_2 H_1$ and $U_2 (P_2 H_1) \subset 
H_{12},$ We obtain that $P_2 H_1 = H_{12}$ and $U_2: H_{12} \to H_{12} ,$ i.e., $H_{12} $ is invariant subspace of the unitary operator 
$U_2.$   

\medskip

{\bf Lemma 3.} {\it The condition (\ref{Z1}) is equivalent to commutativity of projectors:} 
\begin{equation}
\label{Z3a}
[P_1, P_2]=0.
\end{equation}

{\bf Proof.}  a). Let (\ref{Z1}) hold. Then, for any $\psi \in H,$ we have $P_2 P_1 \psi \in H_{12}.$ Hence, $P_1 P_2 P_1 \psi = P_2 P_1 \psi.$
Thus, 
\begin{equation}
\label{Z3}
P_1 P_2 P_1  = P_2 P_1.
\end{equation}
Apply the operation of adjoint operator to the both sides of this equality. We get:    
\begin{equation}
\label{Z4}
P_1 P_2 P_1  = P_1 P_2.
\end{equation}
Hence, we derived (\ref{Z3a}).

b). Let (\ref{Z3a}) hold. Then $P_2 H_1= P_2 P_1 H= P_1 P_2 H\subset H_1$ and, hence, $P_2 H_1 \subset H_{12}.$

\medskip

In the same way the $B-A-B$ condition  implies that     
\begin{equation}
\label{Z1a}
P_1 H_2= H_{12}
\end{equation}
 and 
\begin{equation}
\label{Z2a}
U_1 H_{12} =H_{12}.
\end{equation}

Condition (\ref{Z1a}) is as well equivalent to commutativity of projectors. In particular, the conditions  (\ref{Z1}) and (\ref{Z1a})
are equivalent.  The above results can be presented in the form of Proposition useful for our cognitive applications, see, e.g.,  \cite{BKDB}-\cite{C1}.

\medskip

{\bf Proposition 6.} {\it The combination of the  sequential repeatability  conditions $A-A$ and $B-B$ with  $A-B-A$ or 
$B-A-B$ implies that the projectors commute.} 

\medskip
 
We emphasize that $A-B-A$  implies only $U_2$ invariance of $H_{12}$ and not $U_1$ invariance (as can be seen from the 
Example). In the same way $B-A-B$ implies only $U_1$ invariance of $H_{12}$ and not $U_2$ invariance. Thus, commutativity 
of projectors or conditions (\ref{Z1}) or  (\ref{Z1a}) do not imply $A-B-A$ (nor $B-A-B)$ repeatability.  
 
We can represent $H_j= H_{12} \oplus L_j, j=1,2,$ where $L_j$  is the orthogonal complement of $H_{12}$ 
in the subspace $H_j.$ In other terms $L_j= P_j H \cap (P_1 H \cap P_2 H)^\perp.$ We also can decompose 
 total Hilbert space:    
\begin{equation}
\label{PPrr1}
H=H_{12} \oplus L_j \oplus H_j^\perp.  
\end{equation}
This decomposition will play a crucial role in our further considerations. 
We now show that the sequential repeatability conditions set a rigid constraint onto the geometric inter-relation between
subspaces $L_1$ and $L_2.$


\medskip

{\bf Proposition 6a.} {\it The combination of the  sequential repeatability  conditions $A-A$ and $B-B$ with  $A-B-A$ or 
$B-A-B$ implies that $L_1 \perp L_2.$} 

{\bf Proof.} By Proposition 5 $A-B-A$ sequential repeatability implies that $P_2 H_1 =H_{12}.$ 
We show that, for any pair of  vectors $v \in L_1$ and 
$w \in L_2,$ they are orthogonal. We have $\langle v, w\rangle=\langle v, P_2 w\rangle= \langle P_2v, w\rangle =0,$ because
$P_2v \in H_{12}.$       

\medskip

We remark that in Hilbert space geometry, subspaces $H_1$ and $H_2$ satisfying the condition $L_1 \perp L_2$ are 
called {\it perpendicular}.  It is well known that $P_1, P_2$ are the projectors onto perpendicular subspaces iff they commute.

By Lemma 1 the unitary operators can be represented as block-diagonal operators
\begin{equation}
\label{PPrr3}
U_j=\rm{diag} (V_j, W_j, T_j),  
\end{equation}
where the diagonal blocks are unitary operators acting in spaces $H_{12}, L_j,$
and $H_j^\perp,$ respectively. This invariance decomposition is essential for our further considerations. 
 
\medskip

{\bf Proposition 7.} {\it  The combination of the sequential repeatability conditions $A-A$ and $B-B$ with  
$A-B-A$ and $B-A-B$ is incompatible 
with the order effect.} 

{\bf Proof.} By Proposition 6a the Hilbert space $H$ can be represented as the direct sum:
\begin{equation}
\label{PPrr2}
H=H_{12} \oplus L_1 \oplus L_2 \oplus \tilde{H}, \;  \tilde{H}=   (H_{12} \oplus L_1 \oplus L_2)^\perp.
\end{equation}
Take any $\psi \in H;$ it can be represented as 
$\psi= \psi_{12} + \phi_1 + \phi_2 + \tilde{\psi}$ with components belonging the corresponding subspaces in (\ref{PPrr2}).
First, we calculate $P_1 U_1^* P_2 U_1 P_1 \psi.$ We have $P_1 \psi =  \psi_{12} + \phi_1.$ Then, $U_1 P_1 \psi = 
V_1\psi_{12} + W_1 \phi_1,$ see (\ref{PPrr3}). It is crucial that $V_1: H_{12} \to H_{12},  W_1:  L_1  \to L_1.$ Thus, $W_1 \phi_1 \perp L_2.$
Hence, $P_2U_1 P_1 \psi = V_1\psi_{12}.$ Then we get $U_1^* P_2 U_1 P_1 \psi = V_1^* V_1\psi_{12}= \psi_{12}.$

In the same way we obtain that    $P_2 U_2^* P_1 U_2 P_2 \psi = \psi_{12},$ i.e., no order effect.

 \section{Appendix: infinite-dimensional case}

The following result shows that the considerations of this paper cannot be repeated in the infinite-dimensional case:

\medskip

{\bf Proposition 8.} {\it Let $E=M^*M$. If the dimension of $H$ is infinite, then the equality $E M =M$ does not imply that 
$E$ is an orthogonal projector.}

{\bf Proof.} We construct an example of the operator $E$ which is not an orthogonal projector, but the equality $EM=M$ holds.
Consider an orthonormal basis $(e_j, j=1,2,...)$ in $H.$ Take an arbitrary complex number $a.$ We define the operator $M$ as
\begin{equation}
\label{amba}
  Me_1 = a e_2, \; M e_{n} = e_{n+1}, n \not=1.
\end{equation}
It is easy ro find the adjoint operator:
\begin{equation}
\label{amba1}
M^* e_1 =0,  M^* e_2= \bar{a} e_1, \; M^* e_{n+1}  = e_{n}, n \not=1.
\end{equation}
Then we have
\begin{equation}
\label{amba2}
E  e_1 = M^* Me_1 = a M^* e_2 = \vert a\vert^2 e_1, \; E e_{n}=  M^* M e_{n} = M^* e_{n+1}= e_{n}, n \not=1,
\end{equation}
and 
\begin{equation}
\label{amba3}
E M e_1 = a E e_2 = a e_2  = Me_1, \; E M e_{n} = E e_{n+1}= e_{n+1} =M e_{n}, n \not=1
\end{equation}

\section*{Acknowledgments}

The authors would like to thank J. Busemeyer and E.  Dzhafarov for fruitful discussions and M. D' Ariano, P. Lahti,  W.M. de Muynck, and M. Ozawa 
for knowledge transfer.

\end{document}